 \documentclass[final,5p,times,twocolumn]{elsarticle}

\usepackage{amssymb}
\usepackage{amsmath}

\usepackage{lineno}
\usepackage{hyperref}
\usepackage{subcaption}
\usepackage{upgreek}

\journal{Nuclear Physics B}

\begin{document}

\begin{frontmatter}



\title{Operational experience and performance of the Silicon Vertex Detector after the first long shutdown of Belle II
}


\author[tifr]{K. Ravindran\corref{cor1}}
\ead{krishnakumar.ravindran@tifr.res.in}
\cortext[cor1]{Corresponding author}

\author[warsaw]{K. Adamczyk}
\author[tokyo]{H. Aihara}
\author[warsaw]{S. Bacher}
\author[iitb]{S. Bahinipati}
\author[strasbourg]{J. Baudot}
\author[iitm]{P. K. Behera}
\author[italy1,italy2]{S. Bettarini} 
\author[desy]{T. Bilka}
\author[warsaw]{A. Bozek}
\author[hephy]{F. Buchsteiner}
\author[italy1,italy2]{G. Casarosa} 
\author[tifr]{C. Cheshta}
\author[italy2]{L. Corona} 
\author[mnit]{S. B. Das}
\author[strasbourg]{G. Dujany}
\author[strasbourg]{C. Finck}
\author[italy1,italy2]{F. Forti} 
\author[hephy]{M. Friedl} 
\author[italy2]{A. Gabrielli}
\author[iitb]{V. Gautam}
\author[trieste2]{B. Gobbo}
\author[sokendai,kek]{K. Hara}
\author[ipmu]{T. Higuchi}
\author[hephy]{C. Irmler}
\author[sokendai,kek]{A. Ishikawa}
\author[warsaw]{M. Kaleta}
\author[hephy]{A. B. Kaliyar}
\author[ipmu]{K. H. Kang}
\author[kek]{T. Kohriki}
\author[panjab]{R. Kumar}
\author[mnit]{K. Lalwani}
\author[marseille]{K. Lautenbach}
\author[iitm]{J. Libby}
\author[marseille]{V. Lisovskyi}
\author[italy1,italy2]{L. Massaccesi} 
\author[tifr]{G. B. Mohanty}
\author[italy1,italy2]{S. Mondal} 
\author[sokendai,kek]{K. R. Nakamura}
\author[warsaw]{Z. Natkaniec}
\author[tokyo]{Y. Onuki}
\author[ipmu]{F. Otani}
\author[italy1,italy2]{A. Paladino} 
\author[italy1,italy2]{E. Paoloni} 
\author[iitm]{V. Raj}
\author[tifr]{K. K. Rao}
\author[warsaw]{J.~U.~Rehman}
\author[strasbourg]{I. Ripp-Baudot}
\author[italy1,italy2]{G. Rizzo} 
\author[sokendai]{Y. Sato}
\author[hephy]{C. Schwanda}
\author[marseille]{J. Serrano}
\author[ipmu]{T. Shimasaki}
\author[kek]{J.~Suzuki}
\author[tokyo,kek]{S.~Tanaka}
\author[hephy]{R. Thalmeier}
\author[kek]{T. Tsuboyama}
\author[tokyo]{Y. Uematsu}
\author[trieste1,trieste2]{L. Vitale}
\author[tokyo]{S. J. Wang}
\author[tokyo]{Z. Wang}
\author[warsaw]{J. Wiechczynski}
\author[hephy]{H. Yin}
\author[roma]{L. Zani}
\author[ipmu]{F. Zeng}
\author[]{\\ \vspace{1 mm} (Belle II SVD Collaboration)}
 \affiliation[tifr]{organization={Tata Institute of Fundamental Research},
             city={Mumbai 400005},
             country={India}}

\affiliation[warsaw]{organization={H. Niewodniczanski Institute of Nuclear Physics},
             city={Krakow 31-342},
             country={Poland}}             

\affiliation[tokyo]{organization={Department of Physics, University of Tokyo},
             city={Tokyo 113-0033},
             country={Japan}}

\affiliation[iitb]{organization={Indian Institute of Technology Bhubaneswar},
             city={Bhubaneswar 752050},
             country={India}}             

\affiliation[strasbourg]{organization={IPHC, UMR 7178, Univerisite de Strasbourg, CNRS},
             postcode={67037 Strasbourg},
             country={France}}   
             
\affiliation[iitm]{organization={Indian Institute of Technology Madras},
              city={Chennai 600036},
              country={India}}

\affiliation[italy1]{organization={Dipartimento di Fisica, Università di Pisa},
             postcode={I-56127 Pisa},
             country={Italy}} 
             
\affiliation[italy2]{organization={INFN Sezione di Pisa},
             postcode={I-56127 Pisa},
             country={Italy}}   

\affiliation[desy]{organization={DESY, Deutsches Elektronen Synchrotron},
             postcode={22607 Hamburg},
             country={Germany}}   

\affiliation[hephy]{organization={Institute of High Energy Physics, Austrian Academy of Sciences},
             postcode={1050 Vienna},
             country={Austria}}   

\affiliation[sokendai]{organization={The Graduate University for Advanced Studies (SOKENDAI)},             
             city={Hayama 240-0193}, 
             country={Japan}}

\affiliation[kek]{organization={High Energy Accelerator Research Organization (KEK)},     
             city={Tsukuba 305-0801}, 
             country={Japan}}

\affiliation[ipmu]{organization={Kavli Institute for the Physics and Mathematics of the Universe, University of Tokyo},     
             city={Kashiwa 277-8583}, 
             country={Japan}}

\affiliation[panjab]{organization={Punjab Agricultural University},     
             city={Ludhiana 141004}, 
             country={India}}  

\affiliation[mnit]{organization={Malaviya National Institute of Technology Jaipur},     
             city={Jaipur 302017}, 
             country={India}}  

\affiliation[marseille]{organization={Aix Marseille Universite, CNRS/IN2P3, CPPM},
             postcode={13288 Marseille},
             country={France}}

\affiliation[trieste1]{organization={Dipartimento di Fisica, Università di Trieste},
             postcode={I-34127 Trieste},
             country={Italy}} 
             
\affiliation[trieste2]{organization={INFN Sezione di Trieste},
             postcode={I-34127 Trieste},
             country={Italy}}   

\affiliation[roma]{organization={INFN Sezione di Roma Tre},
             postcode={I-00185 Roma},
             country={Italy}}   



\begin{abstract}
In 2024, the Belle II experiment resumed data taking after the Long Shutdown 1, which was required to install a two-layer pixel detector and upgrade accelerator components.
We describe the challenges of this shutdown and the operational experience thereafter.
With new data, the silicon-strip vertex detector (SVD) confirmed the high hit efficiency, the large signal-to-noise ratio, and the excellent cluster position resolution.  In the coming years, the SuperKEKB peak luminosity is expected to increase to its target value, resulting in a larger SVD occupancy caused by beam background. Considerable efforts have been made to improve SVD reconstruction software by exploiting the excellent SVD hit-time resolution to determine the collision time and reject off-time particle hits. A novel procedure to group SVD hits event-by-event, based on their time, has been developed using the grouping information during reconstruction, significantly reducing the fake rate while preserving the tracking efficiency. The front-end chip (APV25) is operated in the “multi-peak” mode, which reads six samples. A 3/6-mixed acquisition mode, based on the timing precision of the trigger, reduces background occupancy, trigger dead-time, and data size. Studies of the radiation damage show that the SVD performance will not seriously degrade during the detector's lifetime, despite moderate radiation-induced increases in sensor current and strip noise.
\end{abstract}




\end{frontmatter}



\section{Introduction}
\label{sec1}
Belle II~\cite{belle2} is an intensity-frontier experiment designed to record $e^{+}e^{-}$ collision data at the $\Upsilon(4{\rm S})$ resonance, produced by the SuperKEKB asymmetric-energy collider~\cite{superkekb}, which is located at Tsukuba, Japan. SuperKEKB consists of two storage rings, one for 7 GeV electrons and the other for 4 GeV positrons. It is designed to achieve a peak luminosity of $6\times10^{35}~{\rm cm}^{-1}{\rm s}^{-1}$, which corresponds to almost 30 times that of its predecessor, KEKB~\cite{kekb}. So far, Belle II has collected about 575 ${\rm fb}^{-1}$ of physics data and is projected to collect 50~${\rm ab}^{-1}$ of physics data by the end of its lifetime. The data collected by Belle II are used to precisely measure parameters of the Standard Model such as the Cabibbo-Kobayashi-Maskawa matrix elements~\cite{cabibbo,kobayashiandmaskawa}, study $CP$ violation in charm and beauty mesons, perform spectroscopic studies of quarkonium states and exotics, as well as to search for new physics, namely lepton-universality violation in $B$ decays, lepton flavor violation in $\tau$ decays, and light dark matter candidates and others~\cite{belle2book}. 

Belle II started data taking in 2019 and has operated in two principal run periods: Run 1, from March 2019 to June 2022, and Run 2, from January 2024 onward. Long shutdown 1 (LS1) was between these runs.  Both the accelerator and detector were upgraded during LS1. This document presents a comparative analysis of the SVD's performance over the two run periods, provides a detailed account of SVD hardware activities undertaken during LS1, and outlines software developments implemented to maintain consistent performance under high luminosity conditions.

\begin{figure}[h!]
\centering
\includegraphics[width=.9\linewidth]{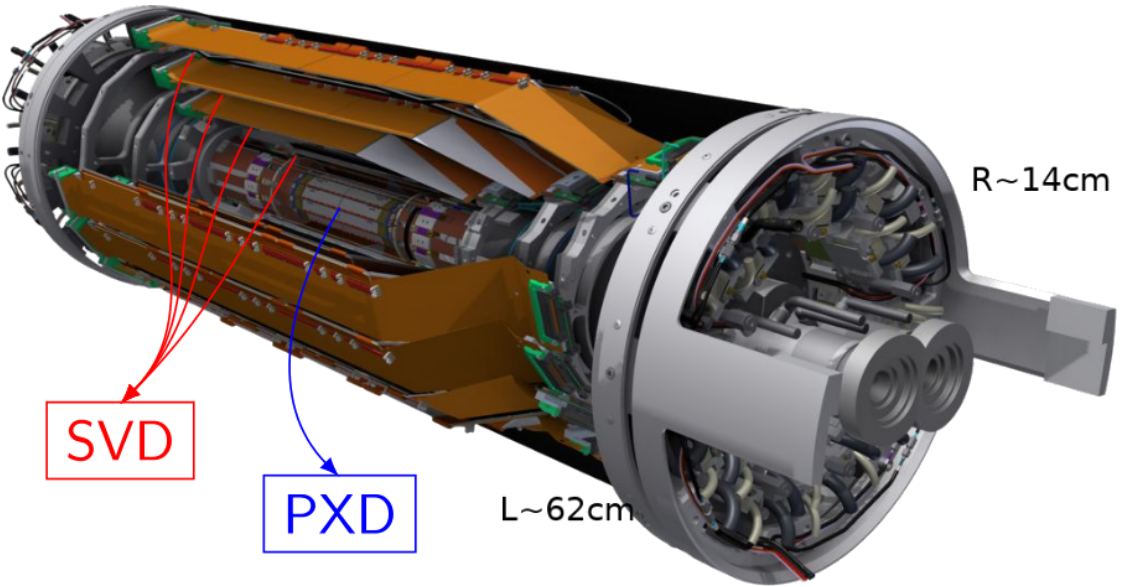}
\caption{Belle II Vertex Detector showing PXD and SVD layers.}\label{fig:svd}
\end{figure}

The Belle II Vertex Detector (VXD), designed for precise vertex reconstruction, is composed of two-layered  Pixel Detectors (PXD)~\cite{pxdpaper} based on Depleted Field Eﬀect Transistor and four-layered SVD~\cite{svdpaper} based on Double-Sided Silicon strip Detector (DSSD). These subdetectors are positioned concentrically around the beam pipe (see Fig.~\ref{fig:svd}), spanning a radial range of 14 to 135 mm and covering the full azimuthal angle. The SVD is developed to perform standalone tracking and identification of low-momentum charged particles, precise vertexing of $K_{S}^{0}$ and $\Lambda^{0}$, and to extrapolate tracks from the Central Drift Chamber to PXD. 

\subsection{Belle II Silicon Vertex Detector}

The SVD layers, numbered 3, 4, 5, and 6 from the inner to outer radius, are composed of structurally and electrically independent DSSD units called ladders. Each DSSD comprises an $n$-type bulk with $p$-doped strips on the $p$-side and $n$-doped strips on the $n$-side, arranged orthogonally to enable precise measurement of the two-dimensional position of a traversing particle. In layers 4-6, the forward-end ladders are slanted relative to the barrel region's cylindrical arrangement to optimize performance while minimizing material and readout channels. The readout strips are AC coupled: the sensing strips are implanted in the $n$-type bulk, and the readout metal strips are placed on top of the implanted strips, separated by a layer of ${\rm SiO}_2$. To achieve a finer effective pitch, a floating strip is incorporated between two readout strips. The fine pitch enables charge sharing between the strips and effectively improves the spatial resolution. The DSSDs, characterized by a high bulk resistivity of 8 k$\Omega$, are produced in three distinct geometries: small rectangular, large rectangular, and trapezoidal. The small rectangular sensors make up the ladders in layer 3, whereas the large rectangular sensors are used in the barrel regions of layers 4-6 and trapezoidal sensors are in the forward regions. The thickness of the rectangular (trapezoidal) sensors is about 320 (300)~$\upmu$m. SVD comprises 172 DSSD sensors, covering an area of $1.2~{\rm m}^{2}$, with a total of 2.2 million readout strips. The full depletion voltage ranges from 20 to 60 V, and the operating voltage is 100 V. The average material budget per layer is 0.7\% radiation length. The radiation hardness of SVD sensors is about 10 Mrad. Radiation detectors based on high-purity diamonds are installed around the beam pipe and SVD ladders to monitor the radiation levels and issue beam-abort signals if the radiation levels exceed a fixed threshold~\cite{diamond}.

The signals from the DSSD are read out by the APV25 ASIC~\cite{apv25} mounted on a flexible circuit board, placed on the top of the sensor, separated by a thin Airex  foam~\cite{airex} for thermal and electrical insulation and to minimize signal cross-talk. This innovative readout design, developed to enhance the signal quality by reducing the capacitive noise, is referred to as the ``Origami" chip-on-sensor design. The signals from the $p$-side are routed to the APV25 via flexible 3-layered polyamide fan-outs wrapped around the sensor edges.  The Origami concept is adopted in all the central sensors, whereas the forward and backward sensors are read out from the sides. 

The power consumption of an APV25 chip is about 0.4 W. To dissipate the heat generated by the chips during operation, dual-phase CO$_2$ cooling lines are placed in direct contact with the chips, with a thermally conductive material sandwiched between them. The APV25 is operated in multi-peak modes; i.e, several analog samples are read out during the shaping time. Given SuperKEKB's current operational phase, with relatively low luminosity and low beam background, six samples are read out by default to reconstruct the output waveform. The 3/6 sample mixed acquisition mode, designed to reduce data readout in high-luminosity conditions, has been developed and successfully tested~\cite{mixeddata}. 

\section{SVD operational experience}
The signals produced in strips are grouped to form clusters that satisfy a fixed signal-to-noise ratio (SNR) threshold. The cluster charge, SNR, and position resolution are used to evaluate the performance of SVD. The techniques to calculate these quantities can be found in Ref.~\cite{svdpaper}.
\subsection{Highlights of performance in Run 1}
During Run 1, the SVD exhibited stable and reliable operation. The fraction of masked readout strips was consistently below 1\%, attributed mainly to manufacturing imperfections. The hit efficiency was greater than 99\% across all the sensors and stable over time. The measured normalized cluster charge of 21,000 $e^{-}$ was compatible with the expected 24,000 $e^{-}$ for the minimum ionizing particles traversing a $320~\upmu$m sensor, accounting for the 15\% uncertainty in the APV25 chip's absolute gain. For most of the runs, the SNR remained within a range of 13-30\%, depending on the sensor position and readout side. A 10-30\% reduction in this value was observed towards the end of the run period due to radiation damage, but it did not compromise the position or hit-time resolution. The position resolution in layer 3 was measured to be 7-12~$\upmu$m on the $p$-side and 15-25~$\upmu$m on the $n$-side, while the cluster-time resolution was 2.9~ns and 2.4~ns for the $p$- and $n$-sides, respectively.
\subsection{Activities during LS1}
During LS1, the VXD underwent a significant upgrade~\cite{ls1}. The existing PXD was replaced with a new two-layered PXD, while the SVD remained unchanged. This 1.5-year upgrade involved a series of intensive hardware activities, including VXD removal, PXD installation, and VXD reinstallation in Belle II. Rigorous test campaigns and self-consistency checks were conducted at each stage to ensure proper functionality. 
The healthiness of all SVD sensors was confirmed during the commissioning of the new VXD in a clean room environment at KEK. Furthermore, alignment and performance studies were conducted before the commencement of Run 2.

\subsection{Performance of SVD in Run 2}

\begin{figure}[h!]
\centering
\includegraphics[width=.8\linewidth]{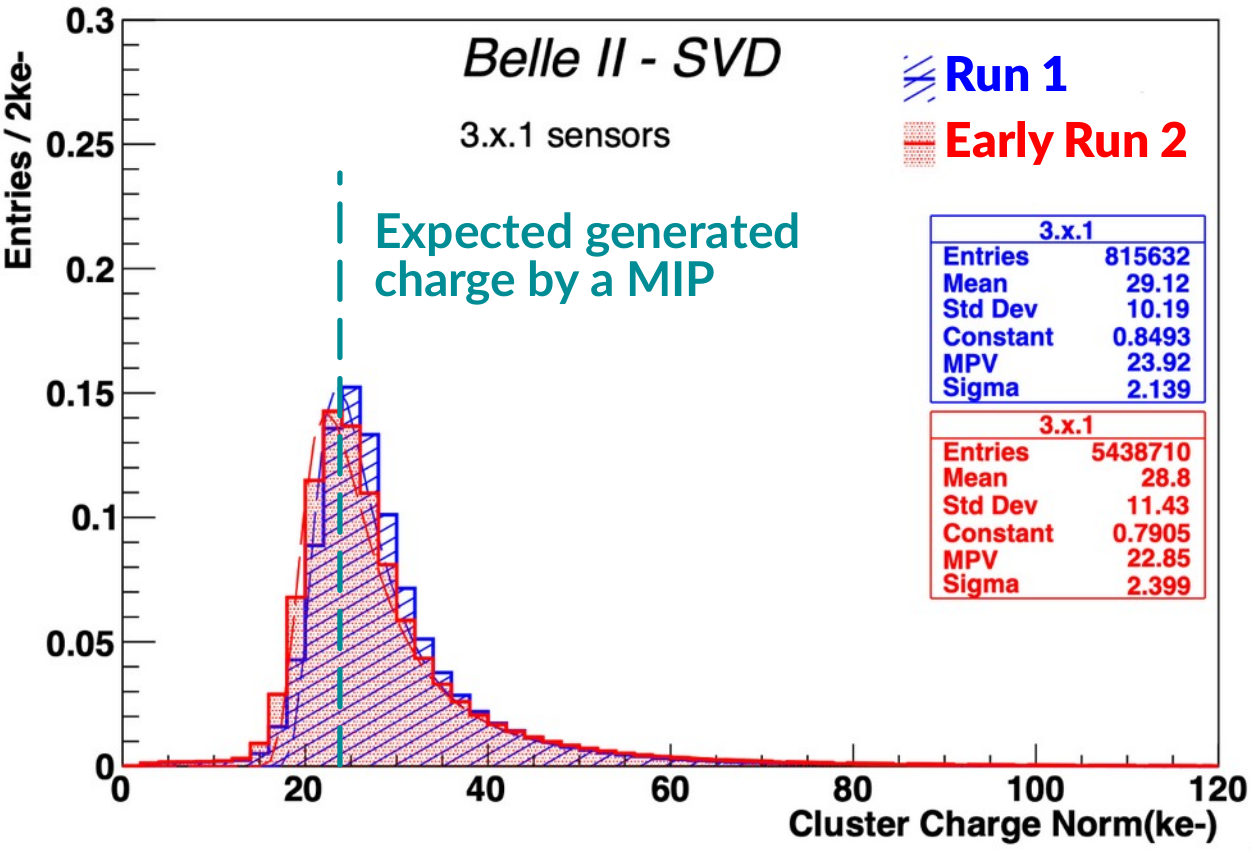}
\caption{Comparison of normalized signal charge collected in Runs~1 and 2.}\label{fig:run1andrun2performance}
\end{figure}

Run 2 presented a different operational environment for the SVD compared to Run 1. The CO$_2$ cooling temperature was lowered to $-25^{~\circ}$C from $-20^{~\circ}$C due to the additional heat load of the completed PXD layer. The average hit occupancy in the innermost layer increased to slightly below 1\%, up from 0.5\% in Run 1, due to the increased beam background in Run 2. However, the operation of SVD was stable and smooth without major problems over the whole period of data taking in Run 2. The evaluation of charge collected, hit efficiencies, and SNR confirmed the same level of performance as Run 1 as illustrated in Fig.~\ref{fig:run1andrun2performance}. 

\begin{figure}[h!]
\centering
\includegraphics[width=.8\linewidth]{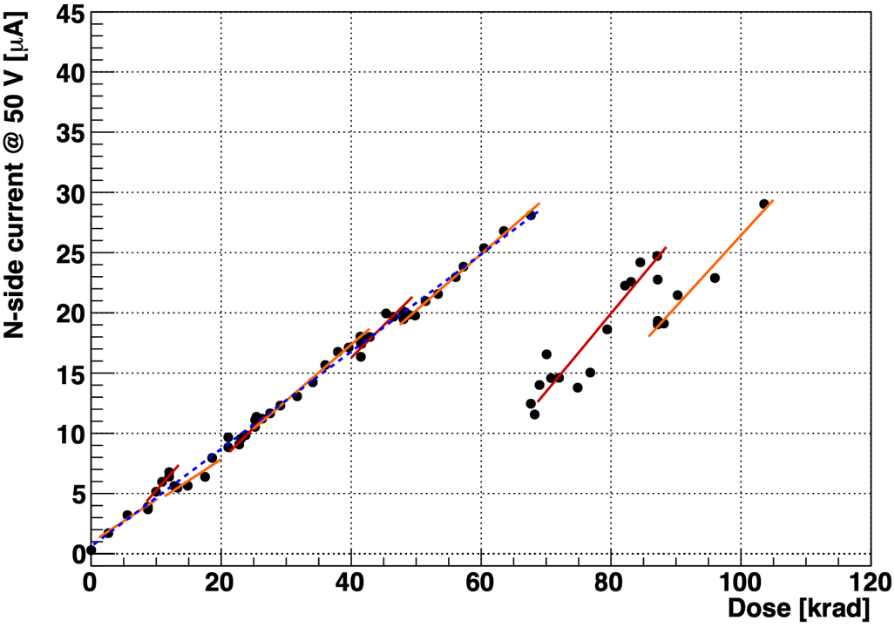}
\caption{Evolution of leakage current as a function of received dose.}\label{fig:radiationplots}
\end{figure}

The total integrated dose on the layer 3 mid-plane, measured by diamond sensors, was approximately 100 krad, corresponding to a fluence of $2.5\times10^{11}$ n$_{\rm eq}$/${\rm cm}^{2}$. Similar to Run~1, the effects of radiation damage on the SVD sensors and its impact on performance were analyzed after collecting data for Run 2. No significant deviation from the expected full depletion voltage was observed across all sensors. On the other hand, a 10\% reduction in the noise level was noted after LS1, attributed to defect annealing and lower operating temperature. The PXD was disabled in May 2024 due to damage to 2\% of its pixels caused by a sudden beam loss. However, the CO$_2$ cooling temperature remained constant, resulting in a lower average noise level throughout Run 2. While multiple sudden beam losses occurred during Run 2, no new pinhole defects were observed in any sensors; the existing pinholes originated either during sensor fabrication or ladder assembly. Although minor, variations in the noise level were observed between rectangular and trapezoidal sensors, likely because of differences in geometry and oxide composition. Additional studies showed that the evolution of leakage current as a function of absorbed dose followed the expected linear trend predicted by the Non-Ionizing Energy Loss hypothesis~\cite{niel}. However, the proportionality constant changed due to the altered temperature settings after the PXD was disabled (see Fig.~\ref{fig:radiationplots}). Based on these studies and the evaluation of operational parameters, we conclude that the radiation damage to the SVD sensors is consistent with the expected dose received, and no significant performance degradation has been observed.

An irradiation campaign was conducted to study the performance of the sensors in the high-luminosity scenario at a test setup of ELPH, Tohoku University, Japan. The sensors were irradiated with a 90 MeV electron beam and a dose of about 10~Mrad $(3\times10^{13}$ n$_{\rm eq}$/${\rm cm}^{2})$, and the noise level and charge collection efficiency were studied. The type inversion was confirmed at 2~Mrad $(6\times10^{12}$ n$_{\rm eq}$/${\rm cm}^{2})$. A type-inverted sensor was conﬁrmed to collect charge well after 10~Mrad of irradiation. This affirms the radiation hardness of silicon sensors used in SVD, even after operating under the full luminosity of SuperKEKB. 

\section{Towards high luminosity}
As discussed earlier, the current hit occupancy is below 1\%, but it is expected to rise as the background increases with higher luminosity in the future. Estimates based on simulation studies predict the value of extrapolated hit occupancy to be around 4.7\% for layer 3 under target luminosity with nominal background conditions. However, the uncertainties due to unknown machine conditions from the planned upgrades~\cite{belle2upgrades} could alter the value significantly. The conservative estimate of hit occupancy stands at 8.7\%. To maintain a consistent performance, various improvements in the reconstruction software have been proposed and studied. 
\begin{figure}[t!]
\centering
\includegraphics[width=.9\linewidth]{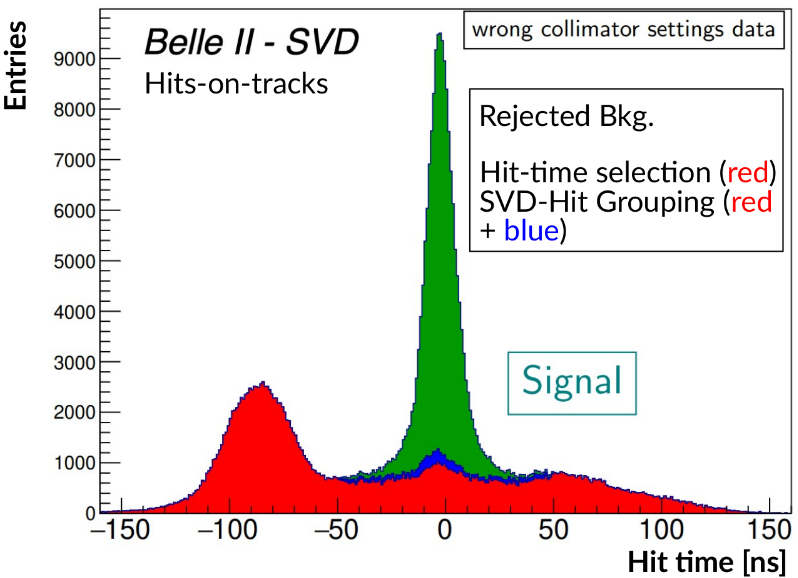} \\ 
\caption{SVD cluster time distribution.}
\label{fig:hittime}
\end{figure}

The SVD exhibited an excellent cluster time resolution ($<$~3~ns), significantly shorter than the approximately 100 ns time range of the beam-induced background. This difference can be leveraged to reject off-time background hits. As illustrated in Fig.~\ref{fig:hittime}, hits from triggered collision products peak at 0~ns, while beam background hits are more or less uniformly distributed with a peak at $-$80 ns corresponding to off-time background hits. With a selection criterion of $|t_{\rm hit}| < 50$~ns and $|t_{\rm hit}^{p} - t_{\rm hit}^{n}| < 20$ ns, where $t_{\rm hit}^{p}$ and $t_{\rm hit}^{n}$ represent the cluster times of $p$-side and $n$-side hits, respectively, about half of the background hits can be rejected while the signal efficiency remains at around 99\%. This method sets the SVD occupancy limit to 4.7\%  without compromising tracking performance. Although tested, these cluster time-based selections have not yet been deployed due to the current low-luminosity operational phase of the experiment.

An alternative approach to background rejection leverages differences in cluster timings by grouping clusters on an event-by-event basis~\cite{grouping}. Since hits originating from the same collision are expected to have similar cluster times, they can be assigned to a common group. This cluster grouping selection method offers significantly better background rejection than the hit time selection method. It further reduces the fake track rate by 15\% while maintaining the same tracking efficiency. 
Track time is defined as the average time of clusters belonging to the track's outgoing arm relative to the collision time. Preliminary studies indicate that incorporating track time information can reduce fake track rates further by a factor of 1.5 in high background environments. Combining cluster grouping with track time-based selections enables setting the layer-3 occupancy limit to approximately 6\% without impacting tracking performance. However, this number is still small compared to the value from a conservative estimate,  opening up discussions about a possible upgrade of VXD~\cite{vtx}.

\section{Conclusions}
Since its initial operation in 2019, the SVD has maintained stable performance and delivered high-quality data, particularly under the challenging background environment of Run~2. The observed radiation damage is consistent with the expectation from the absorbed dose and has not adversely impacted performance. During the LS1, extensive upgrades were performed on the VXD, which included installing a new completed PXD layer within the existing SVD. Several improvements were made to the SVD reconstruction software to ensure robust tracking performance in future high-luminosity scenarios. These include the development of a 3/6-mixed data acquisition mode, as well as new algorithms for hit time-based event selection, cluster grouping, and track time selection. 




\section*{Acknowledgement}
This project has received funding from the European Union’s Horizon 2020 research and
innovation programme under the Marie Sklodowska-Curie grant agreements No 644294, 822070
and 101026516 and ERC grant agreement No 819127. This work is supported by MEXT, WPI
and JSPS (Japan); ARC (Australia); BMBWF (Austria); MSMT (Czechia); CNRS/IN2P3 (France);
AIDA-2020 (Germany); DAE and DST (India); INFN (Italy); and MNiSW (Poland).








\end{document}